\documentclass[12pt]{iopart}

\newtheorem{lemma}{Lemma}
\newtheorem{proposition}{Proposition}
\newtheorem{theorem}{Theorem}
\newtheorem{corollary}{Corollary}
\newtheorem{definition}{Definition}
\newcommand{\be}{\begin{equation}}
\newcommand{\ee}{\end{equation}}

\newcommand{\dif}{\mbox{\rm d}}

\newcommand{\ppsim}{\footnotesize\textcircled{\small$\wedge$}}
\def\Riem{{\rm Riem}}
\def\Ric{{\rm Ric}}

\begin{document}

\title[Characterization of {\rm 2+2} warped spacetimes] {An intrinsic
characterization of 2+2 warped spacetimes}

\author{Joan Josep Ferrando$^1$\
and Juan Antonio S\'aez$^2$}

\address{$^1$\ Departament d'Astronomia i Astrof\'{\i}sica, Universitat
de Val\`encia, E-46100 Burjassot, Val\`encia, Spain.}

\address{$^2$\ Departament de Matem\`atiques per a l'Economia i l'Empresa,
Universitat de Val\`encia, E-46071 Val\`encia, Spain}

\ead{joan.ferrando@uv.es; juan.a.saez@uv.es}

\begin{abstract}
We give several equivalent conditions that characterize the 2+2
warped spacetimes: imposing the existence of a Killing-Yano tensor
$A$ subject to complementary algebraic restrictions; in terms of the
projector $v$ (or of the canonical 2-form $U$) associated with the
2-planes of the warped product. These planes are principal planes of
the Weyl and/or Ricci tensors and can be explicitly obtained from
them. Therefore, we obtain the necessary and sufficient (local)
conditions for a metric tensor to be a 2+2 warped product. These
conditions exclusively involve explicit concomitants of the Riemann
tensor. We present a similar analysis for the conformally 2+2
product spacetimes and give an invariant classification of them. The
warped products correspond to two of these invariant classes. The
more degenerate class is the set of product metrics which are also
studied from an invariant point of view.
\end{abstract}




\section{Introduction}
\label{sec-intro}

The study of warped product metrics is of interest in general
relativity because they include a wide variety of physically
relevant solutions of Einstein equations. For instance, spherically
symmetric spacetimes, Lema\^itre-Friedmann-Robertson-Walker
universes and static metrics are warped products.

Carot and da Costa proposed \cite{Carot-costa} an invariant
characterization of warped product spacetimes and a classification
scheme for them. They also studied the isometry group that each
class can admit. Later, the energy tensor algebraic types compatible
with these classes was analyzed \cite{Haddow-carot}. Recent studies
about warped product spacetimes concern Ricci collineations
\cite{Carot-nunez}, parallel transport \cite{warped-parallel} and
causal structure \cite{Minguzzi}.

Besides the metric tensor, the characterization conditions given in
\cite{Carot-costa} involve vector fields and functions that are not
given in terms of the metric tensor. Consequently, a
characterization of warped spacetimes built with {\em intrinsic}
(depending solely on the metric tensor) and {\em explicit}
conditions is not yet known. The goal of this paper is to obtain
such an intrinsic and explicit characterization.

The intrinsic labeling of metrics starts with the beginning of the
Riemannian geometry \cite{Riemann, Bianchi, Cotton, Weyl, Schouten}.
Cartan \cite{cartan} showed that a metric may be characterized in
terms of the Riemann tensor and its covariant derivatives. Brans
\cite{brans} introduced the Cartan invariant scheme in general
relativity, and Karlhede \cite{karlhede} developed a method to study
the equivalence of two metric tensors. It is worth pointing out that
the covariant determination of the underlying geometry of the Weyl
and Ricci tensors is a necessary tool to characterize spacetimes
intrinsically. This study can be found in \cite{bcm} for the Ricci
tensor and in \cite{fms} for the Weyl tensor.

The intrinsic and explicit characterization of spacetimes is
interesting not only from the conceptual point of view but also from
the practical one. Indeed, it provides an algorithmic way to test if
a metric tensor, given in an arbitrary coordinate system, is a
specific solution of Einstein equations. This is a powerful tool for
subsequent applications. Thus, intrinsic labeling of a solution can
be useful in obtaining a fully algorithmic characterization of the
initial data which correspond to this solution. For instance, the
algorithmic characterization of the Schwarzschild initial data
\cite{GP-VK-2} has been achieved by applying our intrinsic and
explicit labeling of the Schwarzschild geometry \cite{fsS}.
Recently, we have accomplished an analogous study for the Kerr black
hole \cite{fsKerr} which will undoubtedly help to obtain algorithmic
Kerr initial data. Similar approaches, essentially based on the
accurate inquiry of the underlying geometry of the Weyl and Ricci
tensors, have allowed us to intrinsically characterize several
families of spacetimes which are interesting from the physical
and/or geometrical point of view \cite{fms, fsS, fsI, fsWEM, fsKY,
fsIa, fsD, fsWEM2, cfs, fsIb, fs-EM-align, fsEM-sym}.

The conditions that we can find in all the intrinsic
characterizations quoted above are local restrictions on the
curvature tensor. Here we also study local properties and consider
{\em locally warped products}, that is, $n$-dimensional Riemann
spaces satisfying that in a neighborhood of each point there is a
coordinate system $\{x^{\alpha}\}$ such that the metric line element
can be written:
\begin{equation} \label{def-warped}
\dif s^2 = h^1_{ij}(x^k) \dif x^i \dif x^j + e^{2
\lambda(x^k)}h^2_{AB}(x^C) \dif x^A \dif x^B \, ,
\end{equation}
where $i,j,k = 1,..,p$, and $A,B,C = p+1, ..., n$. The two factors
$h^1$ and $h^2$ are $p$ and $n$-$p\,$ dimensional metrics,
respectively, and $\lambda$ is the {\em warping factor}. For short,
we write $g = h^1 + e^{2 \lambda} h^2$ and say that $g$ is a {\em
warped metric}. For a formal definition and basic properties of
warped product metrics see \cite{oneill}.

In relativity we can consider 1+3 and 2+2 warped spacetimes. The 1+3
case can be characterized in terms of the kinematic properties of a
unitary vector field \cite{Carot-costa}. For 2+2 warped spacetimes
the known characterization is not as neat because involves two null
vectors and also the warping factor $\lambda$ \cite{Carot-costa,
Haddow-carot}.

Here we restrict ourselves to study the 2+2 warped spacetimes and we
improve our understanding of them by giving several equivalent
geometric conditions that characterize them: in terms of the
projector $v$ or of the canonical 2-form $U$ associated with the
2-planes of the warped product, or imposing the existence of a
Killing-Yano tensor subject to complementary  properties.

The principal planes, and consequently the invariant tensors $v$ and
$U$, are closely linked with the Ricci and/or Weyl tensors and could
be obtained using the covariant approach given in \cite{bcm, fms}.
This fact enables us to give an intrinsic labeling of the the non
conformally flat 2+2 warped spacetimes imposing conditions on
explicit Weyl concomitants. The results that we present here will
allow us to deal with an old problem stated by Takeno \cite{Takeno}:
the intrinsic an explicit characterization of the spherically
symmetric spacetimes. This question will be analyzed elsewhere
\cite{fsSSST} by considering these metrics as particular 2+2 warped
spacetimes.

A 2+2 warped spacetime is conformal to a 2+2 product spacetime with
particular conformal factor. This fact determines the approach used
here to obtain our results. In Section \ref{sec-product} we analyze
the product metrics: we give several equivalent conditions which
characterize them, we summarize known properties of their curvature
tensor and we offer an algorithmic way to detect them. In Section
\ref{sec-conf2+2} we present a similar study for the conformally 2+2
product metrics and we give an invariant classification of these
spacetimes. In Section \ref{sec-warped} we show that the 2+2 warped
spacetimes correspond to two of these invariant classes, the
t-warped and the s-warped metrics. As a consequence, we obtain the
complementary conditions for a conformally 2+2 product metric to be
a warped spacetime. Moreover we see that, for type D metrics, these
conditions impose restrictions on the Weyl principal 2-form arriving
thus to an intrinsic and explicit characterization of these
spacetimes. The algorithmic nature of our results allows us to
present a flow chart in Section \ref{sec-algo}, which distinguishes
every invariant class of conformally 2+2 product metrics and, in
particular, the t-warped and s-warped spacetimes. Finally, in
Section \ref{sec-discussion} we comment on our main results and work
underway.

In this paper we work on an oriented spacetime with a metric tensor
$g$ of signature $\{-,+,+,+\}$. The Riemann, Ricci and Weyl tensors
are defined as given in \cite{kramer} and denoted, respectively, by
$\Riem(g)$, $R= \Ric(g)$ and $W=W(g)$. For the metric product of two
vectors we write $(x,y) = g(x,y)$, and we put $x^2 = g(x,x)$. Other
basic notation used in this work is summarized in \ref{A-notation}.
In \ref{A-almostproduct} we introduce basic concepts on 2+2
almost-product structures and summarize their invariant
classification. Finally, \ref{A-deterU} presents the covariant
determination of the canonical 2-form of a type D Weyl tensor.

\section{2+2 product spacetimes}
\label{sec-product}

In a 2+2 {\em product spacetime} the metric tensor takes (locally)
the expression $\hat{g} = \hat{v}+ \hat{h}$, where $\hat{v}$ and
$\hat{h}$ denote two 2-dimensional metrics, Lorentzian and
Riemannian, respectively. We have two {\em principal planes}, a
time-like one $V$, with projector $\hat{v}$, and its orthogonal
complement, the space-like plane $H$, with projector $\hat{h}$. The
{\em canonical} 2-{\em form} of the 2+2 product spacetime is the
volume element $\hat{U}$ of the time-like plane, $\hat{U}^2 =
\hat{v}$.

\subsection{Invariant characterization}
\label{subsec-product-a}

In a 2+2 product spacetime the generalized second fundamental forms
$Q_{\hat{v}}$ of $V$ and $Q_{\hat{h}}$ of $H$ vanish, that is, the
principal planes define an integrable and totally geodesic 2+2
almost-product structure. Then, one says that $(\hat{v},\hat{h})$ is
a {\em product structure} and that $\hat{g}$ is a 2+2 {\em product
metric}. In \ref{A-almostproduct} we can find the basic concepts and
properties on almost-product structures. From them we can easily
obtain the following characterizations of the 2+2 product metrics.

\begin{proposition} \label{prop-product-inv}
A spacetime is (locally) a {\rm 2+2} product if, and only if, one of
the following equivalent conditions holds:
\begin{enumerate}
\item There exists a {\rm 2+2} product structure $(\hat{v},\hat{h})$:
$\ Q_{\hat{v}} = 0$, $\ Q_{\hat{h}} = 0\, $.
\item There exists a covariantly constant simple and unitary {\rm
2}-form $\hat{U}$:
\begin{equation}
\tr \hat{U}^2 = 2 \, ,  \qquad \hat{U}\cdot*\hat{U} = 0 \, ,\qquad
\nabla \hat{U} = 0 \, .
\end{equation}
\item There exists a covariantly constant {\rm 2}-projector
$\hat{v}$:
\begin{equation}
\hat{v}^2 = \hat{v} \, ,  \qquad \tr \hat{v} = 2 \, ,\qquad  \nabla
\hat{v} = 0 \, .
\end{equation}
\item There exists a covariantly constant traceless involutive {\rm
2}-tensor $\hat{\Pi}$:
\begin{equation}
\hat{\Pi}^2 = \hat{g} \, ,  \qquad \tr \hat{\Pi} = 0 \, ,\qquad
\nabla \hat{\Pi} = 0 \, .
\end{equation}
\item There exists a simple and unitary Killing-Yano tensor of order two.
\end{enumerate}
\end{proposition}
The 2-form $\hat{U}$ mentioned in (ii) is the canonical 2-form of
the product structure cited in (i), and its square is the projector
$\hat{v}$ cited in (iii). In this statement we can replace $\hat{v}$
by the projector $\hat{h}$ on the space-like plane. The 2-tensor
$\hat{\Pi}$ quoted in (iv) is the structure tensor $\hat{\Pi} =
\hat{v} -\hat{h}$. The Killing-Yano tensor named in (v) is
$\hat{U}$: the Killing-Yano equation implies a vanishing covariant
derivative under the simple and unitary restrictions. Obviously,
$\hat{v} = \hat{U}^2$ is a Killing tensor.

All the characterizations in the proposition above are invariant
properties of the spacetime. Nevertheless, these conditions are not
intrinsic because they involve, besides the metric tensor, other
geometric quantities, namely, the 2-tensors $\hat{U}$, $\hat{v}$ or
$\hat{\Pi}$. In order to obtain a fully intrinsic characterization
we must study the curvature tensor of a product metric.

\subsection{Ricci and Weyl tensors}
\label{subsec-product-b}

The Weyl and Ricci tensor of a product metric have been studied in
detail by many authors. The important fact is that they exclusively
depend on the curvatures of the two 2-dimensional metrics $\hat{v}$
and $\hat{h}$. We summarize some properties that we need here.

\begin{proposition} \label{prop-product-riemann}
Let $\hat{g} = \hat{v} + \hat{h}$ be a {\rm 2+2} product metric, $X$
and $Y$ the Gauss curvature of $\hat{v}$ and $\hat{h}$,
respectively, and $\hat{U}$ such that $\hat{U}^2 = \hat{v}$. It
holds:
\begin{enumerate}
\item $\Riem(\hat{g}) =  - X \ \hat{U} \otimes \hat{U} + Y \,
*\hat{U} \otimes * \hat{U} \, .$
\item $\Ric(\hat{g}) = \Ric(\hat{v})  + \Ric(\hat{h}) = X \,
\hat{v} + Y\, \hat{h} \, .$
\item $\tr \Ric(\hat{g}) = 2( X + Y)$.
\item $W(\hat{g}) = \hat{\rho}(3 \hat{S} + \hat{G}) \, , \quad
\hat{\rho} \equiv - \frac{1}{6} (X+Y) \, , \quad \hat{S} \equiv
\hat{U} \otimes \hat{U} - *\hat{U}\otimes *\hat{U}  \, , \quad
\hat{G} \equiv \frac12 \, \hat{g} \ppsim \hat{g}\, .$
\end{enumerate}
\end{proposition}
As a direct consequence of this proposition we have:
\begin{corollary}
\label{cor-product}
In a {\rm 2+2} product spacetime:
\begin{enumerate}
\item The Weyl tensor is Petrov-Bel type D (or O, i.e. conformally flat)
being $\hat{\rho}= -\frac{1}{6} (X+Y)$ the double eigenvalue. In the
type D case, the Weyl principal planes are the principal planes of
the product.
\item The Ricci tensor is Segre type {\rm [(11)(11)]} (or {\rm [(1111)]},
i.e. Einstein space), with eigenvalues $X$ and $Y$. In the first
case, the Ricci eigen-planes are the principal planes of the
product.
\item
The scalar curvature vanishes if, and only if, the spacetime is conformally
flat.
\end{enumerate}
\end{corollary}

\subsection{Intrinsic and explicit characterization}
\label{subsec-product-c}

Proposition \ref{prop-product-riemann} shows that the 2-form
$\hat{U}$ quoted in point (ii) of proposition \ref{prop-product-inv}
is the principal 2-form of a type D Weyl tensor. Moreover, the
2-tensors $\hat{v}$ and $\hat{\Pi}$ quoted in points (iii) and (iv)
are, respectively, the projector on the time-like eigen-plane of a
Ricci tensor of type [(11)(11)] and the associated structure tensor.
Consequently, these invariant characterizations become intrinsic and
we can state:

\begin{proposition} \label{prop-product-intrinsic-ricci}
A spacetime (which is not an Einstein space) is a {\rm 2+2} product
if, and only if, its Ricci tensor is of type {\rm [(11)(11)]} and
the projector on the Ricci time-like eigen-plane is covariantly
constant.
\end{proposition}

\begin{proposition} \label{prop-product-intrinsic-weyl}
A non conformally flat spacetime is a {\rm 2+2} product if, and only
if, it is Petrov-Bel type D and the principal {\rm 2}-form of the Weyl
tensor is covariantly constant.
\end{proposition}

It is worth remarking that the two propositions above cover all the
metrics with non constant sectional curvature. Moreover, as a
consequence of corollary \ref{cor-product}, a product metric with
constant sectional curvature is, necessarily, flat. Note that a non
flat product metric admits a unique factorization. If the metric is
flat, we have a factorization for every flat plane.

The intrinsic conditions in propositions
\ref{prop-product-intrinsic-ricci} or
\ref{prop-product-intrinsic-weyl} may be tested for a given metric
tensor if we are able to impose the algebraic type and to determine
the Weyl principal 2-form or the Ricci eigen-planes. Moreover, if we
know a covariant expressions of these geometric elements in terms of
the Weyl or Ricci tensors, we can obtain an intrinsic and explicit
labeling of the 2+2 product spacetimes.

Let $\hat{R} \equiv \Ric(\hat{g})$ and $\hat{r} \equiv \tr
\Ric(\hat{g})$ be the Ricci tensor and the scalar curvature of the
product metric $\hat{g}$, and $N$ the Ricci trace-less part,
$N=\hat{R} - \frac{1}{4} \hat{r} \, g$. Then, $\hat{R}$ is of type
[(11)(11)] if, and only if, $\tr N^2 >0$ and $N^2 = \frac{1}{4} \tr
N^2 \, g$. Moreover, the structure tensor $\Pi$ of the Ricci
eigen-planes can be obtained as:
\begin{equation} \label{product-pi}
\Pi = \epsilon \frac{2}{\sqrt{\tr N^2}} \, N  \, , \qquad \epsilon
\equiv \frac{N(x,x)}{|N(x,x)|} \, ,
\end{equation}
where $x$ is an arbitrary time-like vector. Then, under the above
algebraic restrictions, the necessary and sufficient condition for
the metric to be a 2+2 product is $\nabla \Pi=0$. Consequently,
substituting expression (\ref{product-pi}) of $\Pi$, we obtain:

\begin{theorem} \label{theo-product-Ricci}
A spacetime (which is not an Einstein space) is a {\rm 2+2} product if,
and only if, the trace-less part $N$ of the Ricci tensor, $N=\hat{R} -
\frac{1}{4} \hat{r} \, g \not =0$, satisfies:
\begin{equation} \label{product-N}
\tr N^2>0 \, ,\qquad N^2 = \frac{1}{4} \tr N^2 \, g \, , \qquad
\nabla N = \frac{1}{2}\dif \ln [ \tr N^2 ] \, \otimes N \, .
\end{equation}
Moreover the two metric factors can be obtained as $\hat{v} =
\frac12 (g+\Pi)$ and $\hat{h} = \frac12 (g-\Pi)$, where $\Pi$ is
given in {\rm (\ref{product-pi})}.
\end{theorem}

Note that the third (differential) condition in (\ref{product-N}) states
that $N$ is a recurrent tensor, that is, a vector field $e$ exists
such that
\begin{equation} \label{product-recurrent-N}
 \nabla N =  \, e \otimes N \, .
\end{equation}
Moreover, under the two first (algebraic) conditions given in
(\ref{product-N}), equation (\ref{product-recurrent-N}) implies
$e=\frac{1}{2}\dif \ln [ \tr N^2 ]$. Consequently, we can state.
\begin{corollary}
A spacetime (which is not an Einstein space) is a {\rm 2+2} product
if, and only if, the trace-less part $N$ of the Ricci tensor,
$N=\hat{R} - \frac{1}{4} \hat{r} \, g \not =0$, is of type {\rm
[(11)(11)]} and recurrent.
\end{corollary}

We have shown in \cite{fsS} that the Weyl tensor $W$ is type D with
real eigenvalues if, and only if, $\rho \equiv - (\frac{1}{12} \tr
W^3 )^{\frac{1}{3}}\neq 0$ and $\hat{S}^2 + \hat{S} =0$, where
$\hat{S} \equiv \frac{1}{3 \rho} (W - \rho\, G)$. Moreover $\rho$ is
the double Weyl eigenvalue and $S$ depends on the principal 2-form
$\hat{U}$ as $\hat{S} \equiv \hat{U} \otimes \hat{U} -
*\hat{U}\otimes
*\hat{U}$. Then, can be easily shown that $\hat{U}$ is covariantly
constant if, and only if, $\nabla \hat{S} =0$. Consequently,
substituting the expression of $\hat{S}$ in terms of the Weyl tensor
and applying proposition \ref{prop-product-intrinsic-weyl}, we
obtain:

\begin{theorem} \label{theo-product-Weyl}
A non conformally flat spacetime is a {\rm 2+2} product if, and only
if, the Weyl tensor satisfies:
\begin{equation} \label{product-Weyl}
\hspace{-1cm} \rho \equiv - \left(\frac{1}{12} \tr W^3
\right)^{\frac{1}{3}}\neq 0 \, , \quad W^2 + \rho W -2 \rho^2 G = 0
\, , \quad  \nabla W = \dif \ln |\rho| \, \otimes W \, .
\end{equation}
Moreover the two metric factors can be obtained as $\hat{v} =
\hat{U}^2$ and $\hat{h} = \hat{g}-\hat{v}$, where $\hat{U} \equiv
U[W]$ is given in \ref{A-deterU}.
\end{theorem}

Note again that the third (differential) condition in (\ref{product-Weyl})
states that $W$ is a recurrent tensor, that is, a vector field $f$
exists such that
\begin{equation} \label{product-recurrent-Weyl}
 \nabla W =  \, f \otimes W \, .
\end{equation}
Moreover, under the two first (algebraic) conditions given
in (\ref{product-Weyl}), equation (\ref{product-recurrent-Weyl})
implies $f=\dif \ln |\rho|$. Consequently, we can state.
\begin{corollary}
A non conformally flat spacetime is a {\rm 2+2} product if, and only
if, the Weyl tensor is Petrov-Bel type D with real eigenvalues and
recurrent.
\end{corollary}

The intrinsic and explicit characterizations given in theorems
\ref{theo-product-Ricci} and \ref{theo-product-Weyl} allow us to
build an algorithm to test whether a metric tensor is a 2+2 product.
Note that if the spacetime is an Einstein space, the Cotton tensor
vanishes, $\nabla\! \cdot \!W =0$, and then $\nabla U =0$ if, and
only if, the Weyl eigenvalue is constant. Then we arrive to the
algorithm that we present below as a flow chart.

\vspace{-5mm}
\setlength{\unitlength}{0.9cm} {\footnotesize
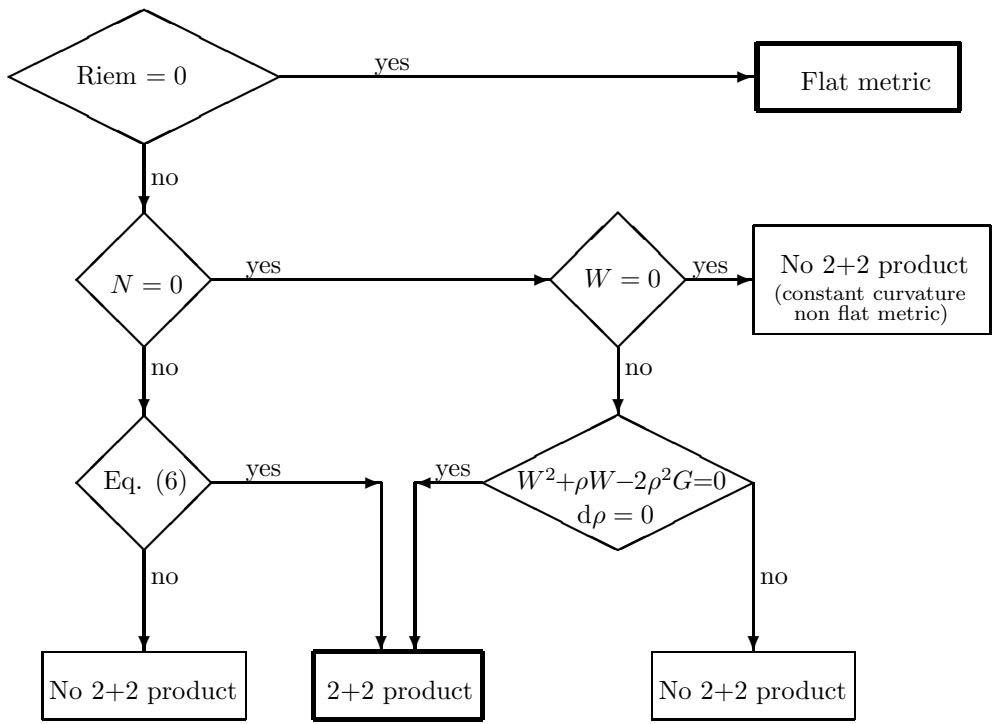
\noindent
\begin{picture}(0,15)
\thicklines

\put(1,13){\line(2,1){2}} \put(1,13){\line(2,-1){2}}
\put(5,13){\line(-2,1){2}}

\put(5,13){\line(-2,-1){2}} \put(2,12.9){Riem $=0$}

\put(5,13){\vector(1,0){7}} \put(6.4,13.1){yes}

\put(3,12){\vector(0,-1){1}} \put(3.1,11.4){no}

{\linethickness{1.5pt} \put(12.05,12.5){\line(0,1){1}}

\put(15.05,12.5){\line(0,1){1}}

\put(12.05,13.5){\line(1,0){3}} \put(12.05,12.5){\line(1,0){3}}}

\put(12.7,12.8){Flat metric}

\put(3,11){\line(-1,-1){1}} \put(3,11){\line(1,-1){1}}
\put(3,9){\line(1,1){1}} \put(3,9){\line(-1,1){1}}
\put(2.5,9.8){$N=0$}

\put(9,10){\line(1,1){1}} \put(9,10){\line(1,-1){1}}
\put(11,10){\line(-1,1){1}} \put(11,10){\line(-1,-1){1}}
 \put(9.5,9.9){$W=0 $}
\put(4,10){\vector(1,0){5}} \put(11,10){\vector(1,0){1}}

{\linethickness{0.5pt} \put(12,9.2){\line(0,1){1.6}}
\put(15.5,9.2){\line(0,1){1.6}} \put(12,9.2){\line(1,0){3.5}}
\put(12,10.8){\line(1,0){3.5}}}

\put(12.4,10.1){No 2+2 product}

\put(12.3,9.7){\scriptsize (constant curvature  }
\put(12.6,9.4){\scriptsize non flat metric)}

\put(10,9){\vector(0,-1){1}} \put(3,9){\vector(0,-1){1}}

\put(3,8){\line(-1,-1){1}} \put(3,8){\line(1,-1){1}}
\put(3,6){\line(1,1){1}} \put(3,6){\line(-1,1){1}} \put(2.4,6.9){Eq.
(\ref{product-N})}

\put(8,7){\line(2,1){2}} \put(8,7){\line(2,-1){2}}
\put(12,7){\line(-2,1){2}}

\put(12,7){\line(-2,-1){2}} \put(8.5,6.9){{$W^2 \hspace*{-1mm}+
\hspace*{-1mm}\rho W \hspace*{-1mm}-\hspace*{-1mm}2 \rho^2 G
\hspace*{-1mm}=\hspace*{-1mm}0$}}

\put(9.4,6.4){$\dif \rho =0 $}



\put(4,7){\vector(1,0){2.5}} \put(3,6){\vector(0,-1){1.5}}
{\linethickness{0.5pt} \put(1.5,3.5){\line(0,1){1}}
\put(4.5,3.5){\line(0,1){1}} \put(1.5,3.5){\line(1,0){3}}
\put(1.5,4.5){\line(1,0){3}}}

{\linethickness{1.5pt} \put(5.5,3.5){\line(0,1){1}}
\put(8,3.5){\line(0,1){1}} \put(5.5,3.5){\line(1,0){2.5}}
\put(5.5,4.5){\line(1,0){2.5}}}

\put(1.6,3.8){No 2+2 product}

\put(5.7,3.8){2+2 product}
 \put(10.6,3.8){No 2+2 product}

{\linethickness{0.5pt} \put(10.5,3.5){\line(0,1){1}}
\put(13.5,3.5){\line(0,1){1}} \put(10.5,3.5){\line(1,0){3}}
\put(10.5,4.5){\line(1,0){3}}}

\put(8,7){\vector(-1,0){1}} \put(7,7){\vector(0,-1){2.5}}
\put(6.5,7){\vector(0,-1){2.5}} \put(12,7){\vector(0,-1){2.5}}

\put(3.1,8.6){no} \put(10.1,8.6){no} \put(4.5,10.1){yes}
\put(11.1,10.1){yes}
 \put(4.5,7.1){yes}
\put(7.3,7.1){yes} \put(3.1,5.5){no} \put(12.1,5.5){no}
\end{picture}

}

\vspace{-3.4cm}

\section{Conformally 2+2 product spacetimes}
\label{sec-conf2+2}

In a {\em conformally} 2+2 {\em product spacetime} the metric tensor
takes the expression $g = e^{2 \lambda } \, \hat{g}$, where
$\hat{g}$ is a (locally) 2+2 product metric, $\hat{g} = \hat{v}+
\hat{h}$. The principal planes $V$ and $H$ have associated the
projectors $v=e^{2 \lambda} \hat{v}$ and $h =e^{2 \lambda} \hat{h}$.
Then $g=v+h$ and $\Pi = v-h$ is the structure tensor. Taking into
account the relations between two conformal metrics, we obtain the
change of the generalized second fundamental forms:
\begin{lemma}
If $g = v+ h =e^{2 \lambda} \, ( \hat{v} + \hat{h} )$, we have
 $$Q_{v} = e^{2 \lambda} [ Q_{\hat{v}} - \hat{v} \otimes
\hat{h}(\dif \lambda)], \qquad Q_{h} = e^{2 \lambda} [ Q_{\hat{h}} -
\hat{h} \otimes \hat{v}(\dif \lambda)] \, .$$
\end{lemma}
Note that this lemma implies the conformal invariance of the
umbilical and integrable properties.

\subsection{Invariant characterization}

If we use the results of the previous section on product metrics and
lemma above, and we take into account that $\hat{v}_{\alpha}^{\beta}
= v_{\alpha}^{\beta}$, we have that the necessary and sufficient
conditions for a metric $g$ to be conformal to a 2+2 product
one is that $g=v+h$ with
\begin{equation} \label{cprod2}
Q_v = - v \otimes h(\dif{\lambda}), \quad Q_h = - h \otimes
v(\dif{\lambda})
\end{equation}
Moreover, $Q_v$ and $Q_h$ determine (and are determined by) the
covariant derivatives of the canonical elements associated with the
structure. Consequently, we obtain the following:
\begin{proposition} \label{prop-cproduct-inv}
A conformally {\rm 2+2} product spacetime is characterized by one of
the following equivalent conditions:
\begin{enumerate}
\item There exists a {\rm 2+2} almost-product structure $(v,h)$ and a
function $\lambda$ such that $\ Q_v = - v \otimes h(\dif \lambda )$,
$Q_h = - h \otimes v(\dif \lambda )\, .$
\item There exists a simple and unitary {\rm 2}-form $U$
($U \! \cdot \! *U=0\, , \ tr U^2 = 2$) such that
\begin{equation} \label{cproduct-invariant-U}
2 \, \nabla U = - U \ppsim a + h \ppsim U(b) \, , \qquad
\dif (a+b)=0\, ,
\end{equation}
where $v \equiv U^2$, $h\equiv g - v$, $a\equiv -*\!U(\nabla \!
\cdot \! *U)$, $b \equiv U(\nabla \! \cdot \! U)$.
\item There exists a traceless and involutive {\rm 2}-tensor
$\Pi$ ($\rm{tr} \Pi=0 \, , \ \Pi^2 =g$) such that
\begin{equation}
2 \nabla \Pi = \Pi \stackrel{{\stackrel{23}{\sim}}}{\otimes} \Phi -
g \stackrel{{\stackrel{23}{\sim}}}{\otimes} \Pi( \Phi )  \, , \quad
\dif \Phi = 0  \,  , \quad \Phi \equiv \frac{1}{2} \Pi (\nabla \!
\cdot \! \Pi) \, .
\end{equation}
\item There exists a simple conformal Killing-Yano tensor of order two.
\end{enumerate}
\end{proposition}
In proposition above, $\ppsim$ and
$\stackrel{{\stackrel{23}{\sim}}}{\otimes}$ denote, respectively,
the anti-symmetrization and symmetrization in the second and third
indexes of the tensorial product (see notation in \ref{A-notation}).

As happens in the 2+2 product case, all these geometrical elements
are closely related to each other. The 2-tensor $\Pi$ cited in (iii)
is the structure tensor of the almost-product structure quoted in
(i), and the 2-form $U$ in (ii) is the associated canonical 2-form.
In (ii) we can substitute the time-like 2-form $U$ by the space-like
one $*U$, $\,2 \, \nabla \! *\!U = - *\!U \ppsim b+ v \ppsim
*\!U(a)\,$. Moreover $\Phi = a+b = -2\, \dif \lambda$. The conformal
Killing-Yano tensor cited in (iv) is $e^{\lambda} U$, and the proof
of this point follows by considering the projections of the
conformal Killing-Yano equation on the planes $V$ and $H$. The
trace-less part of the square of this conformal Killing-Yano tensor
is the conformal Killing tensor $e^{2\lambda}\Pi$.

Carot and Tupper \cite{Carot-Tupper} have classified the conformally
2+2 product spacetimes according to their conformal algebra.
Moreover, they give an invariant characterization which imposes
conditions on two null vectors and also involves the conformal
factor. However, the invariant characterizations given in
proposition \ref{prop-cproduct-inv} involve geometric elements
linked with the curvature tensor. Then, we can obtain an intrinsic
and explicit characterization of a conformally 2+2 product metric if
we know its Ricci and Weyl tensors.

\subsection{Ricci and Weyl tensors}

The Weyl tensor ${W^{\alpha}}_{\beta \mu \nu} $ is conformal
invariant. Then only the Weyl eigenvalue and the Ricci tensor
change. The following proposition summarizes these known results.
\begin{proposition} \label{prop-Ricci-weyl-conf-pro}
\label{prop-conproduct-ricci} Let $g =e^{2 \lambda } \, ( \hat{v} +
\hat{h} )$ be a conformally {\rm 2+2} product metric and let us denote
$\hat{\nabla}$ the covariant derivative associated to the metric
$\hat{g} = \hat{v} + \hat{h}$, and $X$, $Y$ the Gauss curvature of
$\hat{v} $ and $\hat{h}$. Then,
\begin{enumerate}
\item $ \Ric(g) = X \, \hat{v} + Y \, \hat{h} - 2 \left[ \hat{\nabla}
\dif{\lambda}- \dif \lambda \otimes \dif \lambda \right] - \left[
\hat{\Delta} \lambda + 2 \hat{g}(\dif \lambda, \dif \lambda) \right]
\, \hat{g}$
\item The Weyl tensor is type D(or O) with double eigenvalue $\rho
\equiv - \frac{1}{6}e^{-2\lambda} (X+Y)$. In the type D case, the
canonical {\rm 2}-form $U$, $U^2 = v$, is the principal {\rm 2}-form
of the Weyl tensor which takes the expression:
\begin{equation} \label{Weyl-conf-pro}
W = \rho(3 S + G) \, ,  \quad S \equiv U \otimes U -
*U\otimes *U  \, , \quad G \equiv \frac12 \, g
\ppsim g\, .
\end{equation}
\end{enumerate}
\end{proposition}

\subsection{Intrinsic and explicit characterization}

Proposition \ref{prop-conproduct-ricci} shows that the algebraic
type of the Ricci tensor of a conformally 2+2 metric strongly
depends on the Hessian of the conformal factor $\lambda$. Thus, we
have no restrictions on the Segre type and the Ricci tensor is,
generically, algebraically arbitrary.

On the other hand, the Weyl tensor is, necessarily, type D or O. In
the conformally flat case we can only detect the principal structure
if one considers particular conformal factors that restrict the
Ricci tensor type. But we do not consider here these particular
situations. Nevertheless, now we acquire an intrinsic
characterization of the non conformally flat metrics conformal to a
2+2 product because of the 2-form $U$ quoted in point (ii) of
proposition \ref{prop-cproduct-inv} is the Weyl principal 2-form.
\vspace*{-1.5mm}

\begin{proposition} \label{prop-cproduct-intrinsic-weyl}
A non conformally flat spacetime is conformal to a {\rm 2+2} product
if, and only if, it is Petrov-Bel type D and the Weyl principal {\rm
2}-form satisfies {\rm (\ref{cproduct-invariant-U})}.
\end{proposition}

\vspace*{-1.5mm}
The differential conditions (\ref{cproduct-invariant-U}) can be
tested if we know the Weyl principal 2-form $U$. In making more
explicit these conditions we can consider in a first step the Weyl
concomitant $S \equiv \frac{1}{3 \rho} (W - \rho \, G)$ which
depends on $U$ as $S = U \otimes U - *U\otimes *U$. Then, a
straightforward calculation allows us to write
(\ref{cproduct-invariant-U}) in terms of $S$. Moreover, taking into
account the characterization of the type D Weyl tensors with real
eigenvalues \cite{fsS}, one arrives to:
\vspace*{-2mm}

\begin{theorem} \label{teor-conf-product}
A non conformally flat spacetime is conformal to a {\rm 2+2} product
if, and only if, the Weyl tensor satisfies:
\begin{eqnarray} \label{conf-product-S-a}
\rho \equiv - \left(\frac{1}{12} \tr W^3 \right)^{\frac{1}{3}}\neq 0
\, , \quad S^2 + S = 0 \, , \quad \, S \equiv
\frac{1}{3 \rho} (W - \rho\, G) ; \\
\label{conf-product-S-d} 2\, \nabla \! \cdot \! S + 3\, S(\nabla \!
\cdot \! S) -  g \ppsim \Phi = 0 \, , \quad \dif \Phi = 0 \, , \quad
\Phi \equiv \tr[S(\nabla \! \cdot \! S)] \, .
\end{eqnarray}
Moreover the two metric factors can be obtained as $v = U^2$ and $h
= g-v$, where $U \equiv U[W]$ is given in \ref{A-deterU}, and the
conformal factor $\lambda$ is a function satisfying $- 2 \dif
\lambda = \Phi$.
\end{theorem}
\vspace*{-2mm}
The notation used in this theorem is explained in detail in
\pagebreak \ref{A-notation}. Note that $\nabla \! \cdot \!S$ and
$\Phi$ are, respectively, a vector valued 2-form and a 1-form.

Finally, if one introduces the 1-form $\omega = 3\rho(3\rho \Phi - 2
\dif \rho)$ and substitutes the expression of $S$ in terms of the
Weyl tensor in the theorem above, one obtains:

\begin{theorem} \label{teor-conf-product-W}
A non conformally flat spacetime is conformal to a {\rm 2+2} product
if, and only if, the Weyl tensor satisfies:
\begin{eqnarray} \label{conf-product-Weyl}
\rho \equiv - \left(\frac{1}{12} \tr W^3 \right)^{\frac{1}{3}}\neq 0
\, ,
\qquad W^2 + \rho W -2 \rho^2 G = 0 \, ; \\
\label{conf-product-Weyl-b} \hspace{-14mm} \rho \, \nabla \! \cdot
\! W +  W(\nabla \! \cdot \! W) - \frac 13 \, g \ppsim  \omega =0 \,
, \quad \dif \left(\frac{1}{\rho^2}\, \omega\right) =0  \, , \quad
\omega \equiv \tr\left[W(\nabla \! \cdot \! W)\right]\, .
\end{eqnarray}
\end{theorem}

Note that when the Cotton tensor vanishes, $\nabla \! \cdot \! W =
0$, the differential conditions (\ref{conf-product-Weyl-b})
identically hold. Consequently, we recover a result given in
\cite{fsD}: every type D metric with vanishing Cotton tensor is
conformal to a product metric.

\vspace*{-1mm}

\subsection{Invariant classification of conformally {\rm 2+2}
product metrics} \label{subsec-conf2+2-charact}

\vspace*{-1mm}

When a family of spacetimes has an outlined 2+2 almost-product
structure we can classify them by considering the invariant
classification of this structure (see \ref{A-almostproduct}). For
example, elsewhere \cite{fsD} we have classified type D metrics
attending to the differential properties of the Weyl principal
structure, and we have shown that only 16 of the 64 possible classes
are compatible with the vacuum condition.

The principal structure of a conformally 2+2 product spacetime is
umbilical and integrable. Consequently, following the notation given
in \ref{A-notation}, necessarily the structure is of type
$\left(^{00\cdot}_{00\cdot}\right)$, that is, only the classes
$\left(^{001}_{001}\right)$, $\left(^{001}_{000}\right)$,
$\left(^{000}_{001}\right)$ and $\left(^{000}_{000}\right)$ are
possible. For short we will write $\left(^{r}_{s}\right)$ to
indicate the conformally 2+2 product metrics of class
$\left(^{00r}_{00s}\right)$:

\begin{definition}
We say that a conformally {\rm 2+2} product metric is of:
\begin{description}
\item
class $\left(^{0}_{0}\right)$ if both principal planes are minimal
(product metrics).
\item
class $\left(^{0}_{1}\right)$ if the time-like principal plane is minimal
and the space-like one is not.
\item
class $\left(^{1}_{0}\right)$ if the space-like principal plane is minimal
and the time-like one is not.
\item
class $\left(^{1}_{1}\right)$ if none of the principal planes is minimal.
\end{description}
\end{definition}

The classification given in the definition above is based on the
minimal character of the principal planes and imposes differential
conditions on the canonical 2-form $U$. Nevertheless, every class
can also be characterized in terms of the gradient of the conformal
factor. Indeed, as a consequence of proposition
\ref{prop-cproduct-inv}, and taking into account that $-2 \,
\dif\lambda = a+b$, we obtain:

\begin{proposition} \label{prop-classes-cp}
For a conformally {\rm 2+2} product metric $g =e^{2 \lambda } \, (
\hat{v} + \hat{h} )$, let $U$ be the canonical {\rm 2}-form and $a
\equiv -*\!U(\nabla \! \cdot \! *U)$, $b \equiv U(\nabla \!
\cdot \! U)$. Then, the metric is:
\begin{description}
\item
class $\left(^{0}_{0}\right)$ iff $a=b=0$ iff $\lambda$ is constant
($\dif \lambda =0$).
\item
class $\left(^{0}_{1}\right)$ iff $a=0$, $b\not=0$ iff $\dif
\lambda$ lies on the time-like principal plane ($\dif \lambda
\not=0$, $*U(\dif \lambda) =0$).
\item
class $\left(^{1}_{0}\right)$ iff $a\not=0$, $b=0$ iff $\dif
\lambda$ lies on the space-like principal plane ($\dif \lambda
\not=0$, $U(\dif \lambda) =0$).
\item
class $\left(^{1}_{1}\right)$ iff $a\not=0$, $b\not=0$ iff $\dif
\lambda$ does not lie on a principal plane ($U(\dif \lambda)
\not=0$, $*U(\dif \lambda) \not=0$).
\end{description}
\end{proposition}

\vspace*{-1mm}

\section{2+2 warped spacetimes}
\label{sec-warped}

\vspace*{-1mm}

We can consider two different types of 2+2 warped spacetimes
depending on the Lorentzian or Riemannian character of the warped
factor $h_2$ (see expression (\ref{def-warped})). In the first case
the warped plane is time-like, the metric tensor takes the
expression $g = e^{2 \lambda} \hat{v} + h$, with $\hat{v}(\dif
\lambda)=0$, and we say that the spacetime is {\em t-warped}. In the
second case the warped plane is space-like, the metric tensor takes
the expression $g = v + e^{2 \lambda} \hat{h}$, with $\hat{h}(\dif
\lambda)=0$, and we say that the spacetime is {\em s-warped}. Here,
$\hat{v}$ and $v$ denote two 2-dimensional Lorentzian metrics, and
$h$ and $\hat{h}$ denotes two 2-dimensional Riemannian metrics. In
\cite{Haddow-carot} two subclasses of s-warped metrics were
considered depending on the null or non null character of $\dif
\lambda$. But in our study this distinction plays no role.

If for a 2+2 t-warped (respectively, s-warped) metric we define
$\hat{h} =  e^{-2 \lambda}h$ (respectively, $\hat{v} =  e^{-2
\lambda}v$), we obtain that a 2+2 warped metric is conformal to a
2+2 product metric, $g = e^{2 \lambda} (\hat{v} + \hat{h})$, with
$\hat{v}(\dif \lambda)=0$ (respectively, $\hat{h}(\dif \lambda)=0$).
Then, from proposition \ref{prop-classes-cp}, we easily obtain:

\begin{proposition} \label{prop-warped-conformal}
The {\rm 2+2}ç t-warped (respectively, s-warped) spacetimes are the
conformally {\rm 2+2} product spacetimes of class $\left(^{1}_{0}\right)$
(respectively, $\left(^{0}_{1}\right)$).
\end{proposition}

\vspace*{-1mm}

\subsection{Invariant characterization}
\label{subsec-warped-invariant}

\vspace*{-1mm}

As a consequence of proposition \ref{prop-warped-conformal} we can
characterize the warped spacetimes by adding the conditions that
label classes $\left(^{1}_{0}\right)$ or $\left(^{0}_{1}\right)$
(see proposition \ref{prop-classes-cp}) to the conditions given in
section \ref{sec-warped} that characterize the metrics conformal to
2+2 product. Indeed, a simple reasoning leads to:

\begin{proposition} \label{prop-t-warped-inv}
A {\rm 2+2} t-warped spacetime is
characterized by one of the following equivalent conditions:
\begin{enumerate}
\item There exists a {\rm 2+2} almost product structure $(v,h)$ and a
function $\lambda$ such that $\ Q_v = - v \otimes \dif \lambda$,
$Q_h = 0$.
\item There exists a simple, time-like and unitary {\rm 2}-form $U$
($U \! \cdot \! *U=0\, , \ \tr U^2 = 2$) such that
\begin{equation} \label{t-warped-invariant-U}
2 \, \nabla U = - U \ppsim a  \, , \qquad \dif a=0\, , \qquad a\equiv
-*\!U(\nabla \! \cdot \! *U)\not=0 \, .
\end{equation}
\item There exists a time-like {\rm 2}-projector $v$ ($v^2 = v\, ,
\ \tr v = 2 \, , \ v(x,x) < 0 \, ,$ where $x$ in an arbitrary
time-like vector) such that
\begin{equation}
2 \, \nabla v = v \stackrel{{\stackrel{23}{\sim}}}{\otimes} a \,  ,
\qquad \dif a = 0  \,  , \qquad a \equiv  \nabla \! \cdot \! v \not=
0 \, .
\end{equation}
\item There exists a simple and time-like Killing-Yano tensor of order two.
\end{enumerate}
\end{proposition}

\begin{proposition} \label{prop-s-warped-inv}
A {\rm 2+2} s-warped spacetime is
characterized by one of the following equivalent conditions:
\begin{enumerate}
\item There exists a {\rm 2+2} almost product structure $(v,h)$ and a
function $\lambda$ such that $\ Q_v =0$, $Q_h = - h \otimes \dif
\lambda$.
\item There exists a simple, time-like and unitary {\rm 2}-form
$U$ ($U \! \cdot \! *U=0\, , \ \tr U^2 = 2$) such that
\begin{equation} \label{s-warped-invariant-U}
2 \, \nabla\! *\!U = - *\!U \ppsim b  \, , \qquad  \dif b=0\, ,
\qquad b \equiv U(\nabla \! \cdot \! U)\not=0 \, .
\end{equation}
\item There exists a space-like {\rm 2}-projector $h$ ($h^2 = h\, ,
\ \tr h = 2 \, , \ h(x,x) \geq 0 \, ,$ where $x$ in an arbitrary
time-like vector)  such that
\begin{equation}
2 \, \nabla h = h \stackrel{{\stackrel{23}{\sim}}}{\otimes} b  \,  ,
\qquad \dif b = 0  \,  , \qquad b \equiv  \nabla \! \cdot \! h \not=
0\, .
\end{equation}
\item There exists a simple and space-like Killing-Yano tensor of order two.
\end{enumerate}
\end{proposition}

All the geometrical elements in the proposition
\ref{prop-t-warped-inv} (respectively, \ref{prop-s-warped-inv})
above are closely related to each other. The 2-form $U$ of (ii) is
the canonical 2-form of the structure cited in (i), and the 2-tensor
in (iii) is $v= U^2$ (respectively, $h= g-v$). Moreover, $a=-2\dif
\lambda$ (respectively, $b=-2\dif \lambda$). The Killing-Yano tensor
cited in (iv) is $e^{\lambda} U$ (respectively, $e^{\lambda}
*U$). The square of this Killing-Yano tensor is the Killing
tensor $e^{2\lambda}v$ (respectively, $e^{2\lambda}h$).

\vspace*{-1mm}

\subsection{Ricci and Weyl tensors}

\vspace*{-1mm}

Now, to obtain an intrinsic and explicit characterization from the
above invariant properties, we analyze the Ricci and Weyl tensors of
a 2+2 warped metric.

A 2+2 warped spacetime is a conformally 2+2 product with particular
(non constant) conformal factor $\lambda$. Then, the Ricci tensor is
given in proposition \ref{prop-Ricci-weyl-conf-pro}. Now, for a
t-warped (respectively, s-warped) spacetime we have $U(\dif
\lambda)=0$ (respectively, $*U(\dif \lambda)=0$) and, consequently,
$\hat{\nabla} \dif \lambda - \dif \lambda \otimes \dif \lambda $ is
a 2-tensor of the space-like plane $H$ (respectively, time-like
plane $V$). Then, we have the following known result (see, for
example, \cite{Haddow-carot}):
\begin{proposition}
In a {\rm 2+2} t-warped (respectively, s-warped) spacetime the time-like
principal plane (respectively, space-like plane) is an eigen-plane
of the Ricci tensor.
\end{proposition}

When the Ricci eigen-plane quoted in the proposition above
corresponds to a double eigenvalue (Segre types $[(11)1 1]$ or $[1 1
(11)]$), the projectors $v$ or $h$ can be covariantly obtained from
the Ricci tensor and, then, the invariant characterizations given in
point (iii) of propositions \ref{prop-t-warped-inv} or
\ref{prop-s-warped-inv} become intrinsic and explicit. Nevertheless,
there are Segre types admitting a time-like or a space-like
eigen-plane that are more degenerate. In these cases the intrinsic
labeling of the 2+2 warped spacetimes using the Ricci tensor
requires a more detailed analysis that will be considered elsewhere
\cite{fsWarped-Ricci}.

On the other hand, as also stated in proposition
\ref{prop-Ricci-weyl-conf-pro}, the metric is Petrov-Bel type D (or
O). In the type D case the Weyl tensor takes the expression
(\ref{Weyl-conf-pro}) where $\rho \equiv - \frac{1}{6}e^{-2\lambda}
(X+Y)$ is the double eigenvalue and $U$ is the Weyl principal {\rm
2}-form which satisfies $U^2 = v$. Consequently, for non conformally
flat 2+2 warped spacetimes we can obtain an intrinsic labeling.

\vspace{-1mm}

\subsection{Intrinsic and explicit characterization}
\vspace{-1mm}

Propositions \ref{prop-classes-cp} and \ref{prop-warped-conformal}
imply that the 2+2 t-warped (respectively, s-warped) spacetimes are
the conformally 2+2 product ones subject to the complementary
restriction $b=v(\Phi) = 0$ (respectively, $a = h(\Phi) = 0$), a
condition which is equivalent to $U(\Phi) = 0$ (respectively,
$*U(\Phi) = 0$).

On the other hand, for the type D metrics ($W\not=0$) the double
2-form $S$ given in (\ref{Weyl-conf-pro}) satisfies $*S\equiv U
\stackrel{\sim}{\otimes}
*U$, and consequently, $*S(\Phi; \Phi) = U(\Phi) \stackrel{\sim}{\otimes}
*U(\Phi) \, .$
Thus, we arrive to the following:
\begin{proposition} \label{prop-warped-condition}
A (non conformally flat) conformally {\rm 2+2} product is a {\rm
2+2} warped spacetime if, and only if, the Weyl tensor satisfies
$*S(\Phi; \Phi)=0$ where $S \equiv \frac{1}{3 \rho} (W - \rho G)$
and $\Phi \equiv \tr[S(\nabla \! \cdot \! S)] \not= 0$.
\end{proposition}
Note that we have $*W = \rho(3 *\!S -\eta)$, and, consequently, we
can substitute condition $*S(\Phi; \Phi)=0$ in proposition above by
$*W(\Phi; \Phi)=0$.

To distinguish between t-warped and s-warped spacetimes we must add
another condition. Under the assumption $*S(\Phi , \Phi)=0$ we have
that $U(\Phi)=0$ or $*U(\Phi)=0$. Then, the indefinite quadratic
form $Q = S(\Phi; \Phi) = U(\Phi) \otimes U(\Phi)- *U(\Phi) \otimes
*U(\Phi)$ becomes semi-definite: negative for t-warped case and
positive for s-warped spaces. Moreover, the sign of a semi-definite
quadratic form is the sign of its trace with whatever elliptic
metric associated with $g$, $2 x \otimes x +g$, $x$ being a unitary
time-like vector. Thus, we have the following result.
\begin{proposition} \label{prop-t-s-warped-condition}
A non conformally flat {\rm 2+2} warped metric is:
\begin{enumerate}
\item
t-warped if, and only if, $Q = S(\Phi; \Phi)$ is semi-definite
negative, $2 Q(x,x) + \tr Q < 0$.
\item
s-warped if, and only if, $Q = S(\Phi; \Phi)$ is semi-definite
positive, $2 Q(x,x) + \tr Q > 0$.
\end{enumerate}
where $S$ and $\Phi$ are given in proposition {\rm
\ref{prop-warped-condition}} and $x$ is an arbitrary unitary
time-like vector.
\end{proposition}

The two propositions above and theorem \ref{teor-conf-product} that
characterizes the conformally 2+2 product metrics lead to:

\begin{theorem} \label{teor-warped}
For a non conformally flat metric $g$ let $\rho$, $S$ and $\Phi$ be
the Weyl concomitants
\begin{equation} \label{warped-concomitants}
\rho \equiv - \left(\frac{1}{12} \tr W^3 \right)^{\frac{1}{3}}  \neq
0 \, , \quad \, S \equiv \frac{1}{3 \rho} (W - \rho G) \, , \quad
\Phi \equiv \tr[S(\nabla \! \cdot \! S)] \, .
\end{equation}
Then, $g$ is a {\rm 2+2} warped metric if, and only if, it
satisfies:
\begin{equation} \label{warped-conditions}
\hspace{-1.2cm} S^2 + S = 0 \, , \quad 2\, \nabla \! \cdot \! S +
3\, S(\nabla \! \cdot \! S) -  g \ppsim \Phi = 0 \, , \quad \dif
\Phi = 0 \, , \quad *S(\Phi; \Phi)=0 .
\end{equation}
In addition, the metric is t-warped (respectively, s-warped), if,
and only if, $2 S(\Phi,x; \Phi,x) - \Phi^2 < 0$ (respectively, $2
S(\Phi,x; \Phi,x) - \Phi^2 > 0$), where $x$ is an arbitrary unitary
time-like vector.
\end{theorem}

\vspace*{-5mm}

\section{A summary in algorithmic form}
\label{sec-algo}

\vspace*{-2mm}

Theorems \ref{teor-conf-product} and \ref{teor-warped} unable us to
perform an algorithm which distinguishes, between all the non
conformally flat metrics, the four classes of conformally 2+2
product spacetimes: $\left(^{0}_{0}\right)$ (product),
$\left(^{1}_{0}\right)$ (t-warped), $\left(^{0}_{1}\right)$
(s-warped) and $\left(^{1}_{1}\right)$ (regular conformally
product). The flow chart (see below) shows the role played by every
condition in these theorems. The input data are the Weyl
concomitants $\rho=\rho(g)$, $S=S(g)$ and $\Phi=\Phi(g)$ defined in
(\ref{warped-concomitants}).

\vspace*{1.8cm}

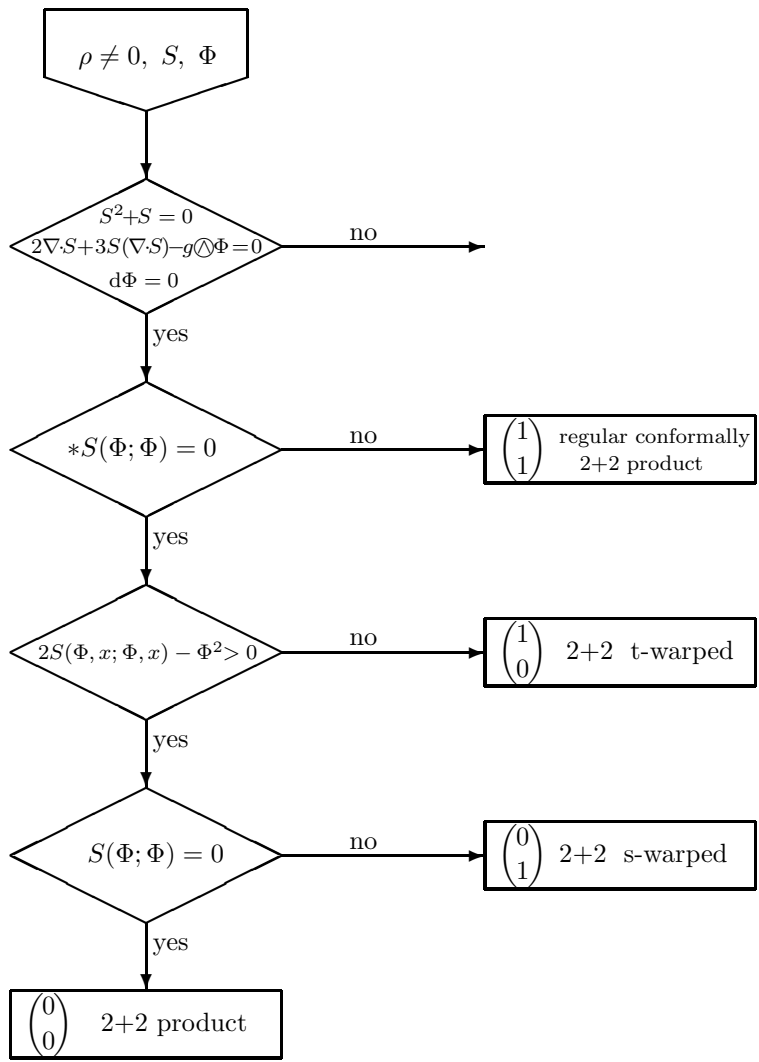
\setlength{\unitlength}{0.9cm} {\footnotesize \noindent

\begin{picture}(0,18)
\thicklines

\put(3,18){\line(3,1){1.5}}

\put(1.5,18.5){\line(3,-1){1.5}}

\put(1.5,18.5){\line(0,1){1}}\put(4.5,19.5){\line(-1,0){3}}
\put(4.5,19.5){\line(0,-1){1}} \put(2,18.7){$  \rho \neq 0  , \ S ,
\ \Phi$} \put(3,18){\vector(0,-1){1}}

\put(1,16){\line(2,1){2}}\put(1,16){\line(2,-1){2}}
\put(5,16){\line(-2,1){2}} \put(5,16){\line(-2,-1){2}}

\put(3,15){\vector(0,-1){1}}

\put(1,13){\line(2,1){2}} \put(1,13){\line(2,-1){2}}
\put(5,13){\line(-2,1){2}} \put(3,12){\vector(0,-1){1}}
\put(5,13){\line(-2,-1){2}}

\put(1,10){\line(2,1){2}} \put(1,10){\line(2,-1){2}}
\put(5,10){\line(-2,1){2}} \put(5,10){\line(-2,-1){2}}

\put(3,9){\vector(0,-1){1}}

\put(3,8){\line(-2,-1){2}} \put(3,8){\line(2,-1){2}}
\put(3,6){\line(2,1){2}} \put(3,6){\line(-2,1){2}}

\put(3,6){\vector(0,-1){1}}

\put(5,13){\vector(1,0){3}} \put(5,10){\vector(1,0){3}}
\put(5,7){\vector(1,0){3}}

\put(5,16){\vector(1,0){3}}

\put(1,4){\line(1,0){4}} \put(1,4){\line(0,1){1}}
\put(5,5){\line(-1,0){4}} \put(5,5){\line(0,-1){1}}
\put(1.2,4.4){$\left( \hspace*{-2mm} \begin{array}{l} 0 \\0
\end{array} \hspace*{-2mm} \right) $}
\put(2.2,4.4){ {2+2 product}}

\put(8,6.5){\line(1,0){4}} \put(8,6.5){\line(0,1){1}}
\put(12,7.5){\line(-1,0){4}} \put(12,7.5){\line(0,-1){1}}
\put(8.2,6.9){$\left( \hspace*{-2mm} \begin{array}{l} 0\\1
\end{array} \hspace*{-2mm} \right) $}
\put(9.1,6.9){2+2 \  s-warped}

\put(8,9.5){\line(1,0){4}} \put(8,9.5){\line(0,1){1}}
\put(12,10.5){\line(-1,0){4}} \put(12,10.5){\line(0,-1){1}}
\put(8.2,9.9){$\left( \hspace*{-2mm} \begin{array}{l} 1\\0
\end{array} \hspace*{-2mm} \right) $}
\put(9.2,9.9){2+2 \  t-warped}

\put(8,12.5){\line(1,0){4}} \put(8,12.5){\line(0,1){1}}
\put(12,13.5){\line(-1,0){4}} \put(12,13.5){\line(0,-1){1}}
\put(8.2,12.9){$\left( \hspace*{-2mm} \begin{array}{l} 1\\1
\end{array} \hspace*{-2mm} \right) $}
\put(9.4,12.7){\scriptsize 2+2 product}
\put(9.1,13.1){\scriptsize regular conformally}

\put(1.3,15.9){\scriptsize$2 \nabla \hspace*{-1mm} \cdot
\hspace*{-1mm} S \hspace*{-0.7mm}+ \hspace*{-0.6mm} 3 S(
\hspace*{-0.2mm}\nabla\hspace*{-1mm} \cdot \hspace*{-1mm}
S\hspace*{-0.2mm}) \hspace*{-1mm}- \hspace*{-1mm}g \hspace*{-0.3mm}
\wedge \hspace*{-3.2mm}   \bigcirc \hspace*{-0.3mm}\Phi
\hspace*{-0.5mm}  = \hspace*{-0.6mm}0 $}

\put(2.3,16.35){\scriptsize $S^2  \hspace*{-1mm} +  \hspace*{-1mm}S
= 0 $}

\put(2.45,15.4){\scriptsize $\dif \Phi = 0 $}

\put(1.7,12.9){ $*S(\Phi ; \Phi)=0$}

\put(1.4,9.9){\scriptsize $ 2 S(\Phi, x; \Phi, x) - \Phi^2
\hspace*{-1mm}> 0$}

\put(2,6.9){ $ S(\Phi;\Phi) = 0 $}

\put(6,7.1){\footnotesize no} \put(6,10.1){\footnotesize no}

\put(6,13.1){\footnotesize no} \put(6,16.1){\footnotesize no}

\put(3.1,5.6){\footnotesize yes} \put(3.1,8.6){\footnotesize yes}
\put(3.1,11.6){\footnotesize yes} \put(3.1,14.6){\footnotesize yes}
\end{picture}
}

\vspace{-3.5cm}

\section{Discussion and work in progress}
\label{sec-discussion}

In this work we have acquired an intrinsic and explicit labeling of
the non conformally flat 2+2 warped spacetimes (theorem
\ref{teor-warped}). This result is based on the invariant
characterization of these spacetimes in terms of a time-like and
unitary 2-form $U$ which result to be the canonical 2-form of the
(type D) Weyl tensor. In contrast, the invariant characterization
given previously \cite{Carot-costa, Haddow-carot} imposes conditions
on two null vectors and it also involves the warping factor
$\lambda$.

In \cite{Haddow-carot} the warped spacetimes were classified taking
into account the projection of the gradient of the warped factor on
two null directions. Here we have shown that t-warped and s-warped
spacetimes correspond to two classes of conformally product metrics
when one applies a general classification of the 2+2 almost-product
structures.

In the non conformally flat case we have labeled with intrinsic and
explicit conditions (theorems \ref{theo-product-Weyl},
\ref{teor-conf-product} and \ref{teor-warped}) the four compatible
classes of this classification (product, t-warped, s-warped and
regular conformally product). Thus, we were able to build an
algorithm to distinguish them that we have presented as a flow
chart.

Conformally flat product metrics have been fully studied here
(theorem \ref{theo-product-Ricci}). Nevertheless, the intrinsic
characterization of the 2+2 warped conformally flat spacetimes
requires a more detailed analysis of the different Ricci algebraic
types that will be considered elsewhere \cite{fsWarped-Ricci}.

We have also shown in this work that the four classes quoted above
can also be characterized in terms of a Killing-Yano or a conformal
Killing-Yano tensor (see propositions \ref{prop-product-inv},
\ref{prop-cproduct-inv}, \ref{prop-t-warped-inv} and
\ref{prop-s-warped-inv}). The existence of quadratic first integrals
of the geodesic equation in these spacetimes has been shown in
\cite{Haddow-carot}. But here we have shown that, the existence of
an algebraically special Killing-Yano or conformal Killing
Yano-tensor is a necessary and sufficient condition for a metric to
be of a fixed class.

It is worth remarking that if a metric tensor, given in arbitrary
coordinates, satisfies our characterization theorems, from the
Riemann tensor we can obtain the geometric elements that appear in
its canonical form, namely, the two factors $v$ and $h$ and the
conformal (or warping) factor $\lambda$ (see theorem
\ref{teor-conf-product}).

Finally, we wish to comment that our results on s-warped metrics
pave the way to acquiring the intrinsic labeling of the spherically
symmetric spacetimes, a question that is considered elsewhere
\cite{fsSSST}.

\ack We would like to thank J. Carot and J. A. Morales-Lladosa for
their comments. This work has been supported by the Spanish
Ministerio de Ciencia e Innovaci\'on, MICIN-FEDER project
FIS2009-07705.

\appendix

\section{Notation}
\label{A-notation}

\begin{enumerate}
\item
{\bf Products and other formulas involving 2-tensors $A$ and $B$}:
\begin{enumerate}
\item
Composition as endomorphisms: $A \cdot B$,
\begin{equation}
(A \cdot B)^{\alpha}_{\ \beta} = A^{\alpha}_{\ \mu} B^{\mu}_{\
\beta}
\end{equation}
\item
Square and trace as an endomorphism:
\begin{equation}
A^2 = A \cdot A \, , \qquad \tr A = A^{\alpha}_{\ \alpha}.
\end{equation}
\item
Action on a vector $x$, as an endomorphism $A(x)$, and as a
quadratic form $ A(x,x)$:
\begin{equation}
A(x)^{\alpha} = A^{\alpha}_{\ \beta} x^{\beta}\, , \qquad A(x,x) =
A_{\alpha \beta} x^{\alpha} x^{\beta} \, .
\end{equation}
\item
Exterior product as double 1-forms: $A \ppsim B$ is the double
2-form,
\begin{equation}
(A \ppsim B)_{\alpha \beta \mu \nu} = A_{\alpha \mu} B_{\beta \nu} +
A_{\beta \nu} B_{\alpha \mu} - A_{\alpha \nu} B_{\beta \mu} -
A_{\beta \mu} B_{\alpha \nu} \, .
\end{equation}
\item
Exterior product with a vector $x$ as double 1-forms: $A \ppsim x$
is the vector-valued 2-form,
\begin{equation}
(A \ppsim x)_{\alpha,\, \mu \nu} = A_{\alpha \mu} x_{\nu} -
A_{\alpha \nu} x_{\mu}  \, .
\end{equation}
\item
Symmetrized tensorial product with a vector $x$ as double 1-form: $A
\stackrel{{\stackrel{23}{\sim}}}{\otimes} x$ is the vector-valued
symmetric 2-tensor,
\begin{equation}
(A \stackrel{{\stackrel{23}{\sim}}}{\otimes} x)_{\alpha,\, \mu \nu}
= A_{\alpha \mu} x_{\nu} + A_{\alpha \nu} x_{\mu}  \, .
\end{equation}
\end{enumerate}
\item
{\bf Products and other formulas involving double 2-forms $P$ and
$Q$}:
\begin{enumerate}
\item
Composition as endomorphisms of the 2-forms space: $P \circ Q$,
\begin{equation}
(P \circ Q)^{\alpha \beta}_{\ \ \rho \sigma} = \frac12 P^{\alpha
\beta}_{\ \ \mu \nu} Q^{\mu \nu}_{\ \ \rho \sigma}
\end{equation}
\item
Square and trace as an endomorphism:
\begin{equation}
P^2 = P \circ P \, , \qquad \Tr P = \frac12 P^{\alpha \beta}_{\ \
\alpha \beta}.
\end{equation}
\item
Action on a 2-form $X$, as an endomorphism $P(X)$, and as a
quadratic form $P(X,X)$,
\begin{equation}
P(X)_{\alpha \beta} = \frac12 P_{\alpha \beta}^{\ \ \mu \nu} X_{\mu
\nu} \, , \qquad P(X,X) = \frac14 P^{\alpha \beta \mu \nu} X_{\alpha
\beta} X_{\mu \nu}.
\end{equation}
\item
The Hodge dual operator is defined as the action of the, metric
volume element $\eta$ on a 2-form $F$ and a double 2-form $W$:
\begin{equation}
*F = \eta(F)  \, , \qquad  *W = \eta \circ W \, .
\end{equation}
\item
Action on two vectors $x$ and $y$, $P(x;y)$,
\begin{equation}
P(x;y)_{\alpha \beta} = P_{\alpha \mu \beta \nu} x^{\mu} y^{\nu} \,
.
\end{equation}
\item
Action on a vector-valued 2-form $Y$ as an endomorphism $P(Y)$,
\begin{equation}
P(Y)_{\lambda, \, \alpha \beta} = \frac12 P_{\alpha \beta}^{\ \ \mu
\nu} Y_{\lambda, \, \mu \nu} \, .
\end{equation}
\item
The trace of a vector-valued 2-form $Y$ is the 1-form $\tr Y$,
\begin{equation}
(\tr Y)_{\alpha} = g^{\lambda \mu} Y_{\lambda,\, \mu \alpha} \, .
\end{equation}
\end{enumerate}
\end{enumerate}

\section{2+2 almost-product structures}
\label{A-almostproduct}

The generalized second fundamental form $Q_v$ of a non-null p--plane
$V$ (set of vector fields generated by p independent vector fields)
is the (2,1)-tensor:
\begin{equation}
Q_v(x,y) = h(\nabla_{v(x)}v(y)) \, , \qquad \forall \; \; x,y
\end{equation}
where $v$ is the projector on $V$ and $h= g-v$. Let us consider the
invariant decomposition of $Q_v$ into its antisymmetric part $A_v$
and its symmetric part $S_v \equiv S_v^T + {1 \over p}v \otimes \tr
S_v$, where $S_v^T$ is a traceless tensor:
\begin{equation}  \label{Q2}
Q_v = A_v + \frac{1}{p} v \otimes \tr S_v + S_v^T
\end{equation}
The p-plane $V$ is a {\em foliation} if, and only if, $A_v =0$, and,
similarly, $V$ is said to be {\em minimal, umbilical or geodesic} if
$\tr S_v=0$, $S_v^T =0$ or $S_v  =0$, respectively.

On the spacetime, a 2+2 almost-product structure is defined by a
time-like plane $V$ and its space-like orthogonal complement $H$.
Let $v$ and $h= g-v$ be the respective projectors and let $\Pi =
v-h$ be the {\it structure tensor}. A 2+2 spacetime structure is
also determined by the {\it canonical} unitary 2-form $U$, volume
element of the time-like plane $V$. Then, the respective projectors
are $v=U^2$ and $h = -(*U)^2$.

The 2+2 almost-product structures can be classified by taking into
account the invariant decomposition of the covariant derivative of
the structure tensor $\Pi$ or, equivalently, according to the
foliation, minimal or umbilical character of each plane
\cite{fsD,cfs}. Every principal 2--plane can be subject or not to
three properties, so $2^6 = 64$ classes can be considered.
\begin{definition}
Taking into account the foliation, minimal or umbilical character of
each $2$-plane we distinguish $64$ different classes of {\rm 2+2}
almost-product structures.

We denote the classes as $\left(^{p \, q \,r}_{lmn}\right)$, where the
superscripts $p,q,r$ take the value $0$ if the time-like principal
plane is, respectively, a foliation, a minimal or an umbilical
plane, and they take the value $1$ otherwise. In the same
way, the subscripts $l, m , n$ collect the foliation, minimal or
umbilical nature of the space-like plane.
\end{definition}

The most degenerated class that we can consider is
$\left(^{000}_{000}\right)$ which corresponds to a product
structure, and the most regular one is $\left(^{111}_{111}\right)$
which means that neither $V$ nor $H$ are foliation, minimal or
umbilical planes. We will put a dot in place of a fixed
script (1 or 0) to indicate the set of metrics that cover both
possibilities. So, for example, the metrics of type
$\left(^{111}_{11 \, \cdot}\right)$ are the union of the classes
$\left(^{111}_{111}\right)$ and $\left(^{111}_{110}\right)$; or a
metric is of type $\left(^{0 \, \cdot \, \cdot}_{\, \cdot \, \cdot
\, \cdot}\right)$ if the time-like 2--plane is a foliation.

We will say that a structure is integrable when both, $V$ and $H$
are a foliation, that is, of type $\left(^{0 \, \cdot \, \cdot}_{0
\, \cdot \, \cdot}\right)$. We will say that the structure is
minimal (respectively, umbilical) if both, $V$ and $H$ are minimal
(respectively, umbilical), that is, of type $\left(^{ \cdot \ 0 \,
\cdot}_{ \cdot \ 0 \, \cdot}\right)$ (respectively, $\left(^{\cdot
\, \cdot \ 0}_{\cdot \, \cdot \ 0}\right)$).

\section{Covariant determination of the canonical 2-form $U$ of a type D
Weyl tensor with real eigenvalues} \label{A-deterU}

The covariant determination of the canonical elements of the Weyl
tensor for all the Petrov-Bel types was presented in \cite{fms}
using the self-dual formalism. In a more recent paper \cite{fsKerr}
we have also given the expression of the canonical 2-form $U$ of a
type D Weyl tensor using real formalism. Here we particularize it
for the case of real eigenvalues:
\begin{lemma} \label{lemma-determ-U-real}
For a Petrov-Bel type D Weyl tensor with real eigenvalue $\rho
\equiv - (\frac{1}{12} \tr W^3 )^{\frac{1}{3}}$, the canonical {\rm
2}-form $U$ can be obtained as:
\begin{equation} \label{determ-U-real}
U = U[W] \equiv \frac{1}{\chi \sqrt{\chi + f}}\left((\chi + f) \, F
+ \tilde{f} \, *F\right) \, ; \qquad F \equiv P(Z) \, ,
\end{equation}
where $Z$ is an arbitrary {\rm 2}-form and
\begin{equation}
\hspace{-1cm} P \equiv W - \rho \, G \, , \quad \chi \equiv
\sqrt{f^2 + \tilde{f}^2} \, , \quad f \equiv \tr F^2 \, , \quad
\tilde{f} \equiv \tr (F \! \cdot \! *F) \, . \label{determ-U-real-2}
\end{equation}
\end{lemma}

\end{document}